\documentclass{ws-procs975x65}

\begin{document}

\title{Mirror dark matter\footnote{Talk given in the Festschrift in honour of
G. C. Joshi and B. H. J. McKellar, November 2006.}}

\author{R. Foot$^{**}$}

\address{School of Physics,University of Melbourne,Victoria 3010 Australia\\
$**$ E-mail: rfoot@unimelb.edu.au}

\begin{abstract}
A mirror sector of particles and forces provides a simple
explanation of the inferred dark matter of the Universe. 
The status of this theory is reviewed - with emphasis on how the theory 
explains the impressive DAMA/NaI annual modulation signal, whilst also
being consistent with the null results of the other
direct detection experiments.
\end{abstract}

\vskip 0.7cm

\bodymatter

There is strong evidence for non-baryonic dark matter in the Universe from
observations of flat rotation curves in spiral galaxies, from precision
measurements of the CMB and from the DAMA/NaI annual modulation signal. 
The standard model of particle physics has no candidate particles. 
Therefore new 
particle physics is suggested. 

There are four most basic requirements for a dark matter candidate:
\begin{itemize}
\item Massive - The elementary particle(s) comprising the 
non-baryonic dark matter need to have mass.

\item Dark - The dark matter particles couple very weakly to ordinary photons 
(e.g. electrically neutral particles).

\item Stable - The lifetime should be greater than about 10 billion years.

\item Abundance - $\Omega_{dark} \approx 5\Omega_b$ (inferred from WMAP CMB observations\cite{wmap}).

\end{itemize}

It is not so easy to get suitable candidates from particle physics satisfying these four
basic requirements. A popular solution is to hypothesize new neutral particles which are weakly
interacting (WIMPs), but this doesn't necessarily make them stable. In fact, the most natural life-time
of a hypothetically weakly interacting particle is very short:
\begin{eqnarray}
\tau (wimp) &\sim & \frac{M_W^4}{g^4 M^5_{wimp} } 
  \sim  10^{-24}\ {\rm seconds \ - \ if } \ M_{wimp} \sim M_Z\ .
 \end{eqnarray}
 This is about 41 orders of magnitude too short lived! Of course there is a trivial solution - which is to invent a symmetry 
 to kinematically forbid the particle to decay, but this is ugly because it is ad hoc. The proton and electron, for example, 
 are not stabalized
 by any such ad hoc symmetry\footnote{Protons and electrons are stabalized by baryon and lepton number
 $U(1)$ global symmetries which are not imposed, but are accidental symmetries of the standard model. These symmetries cannot be broken by any renormalizable term consistent with the gauge symmetries in the standard model.}. 
 It is reasonable to suppose that the dark matter particles, like the proton and electron, 
 will also have a good reason for their stability.
 On the other hand, we also know that the standard model works very well. There is no evidence for anything new (except
 for neutrino masses). For example, precision electroweak tests are all nicely consistent with no new
 physics. 
 
 A simple way to introduce dark matter candidates which are naturally dark, stable, massive 
 and don't modify standard model physics is to introduce a mirror sector of particles and forces\cite{flv}.
 For every standard model particle there exists a mirror partner\footnote{For a more comprehensive review,
see. e.g. ref.\cite{review} .}, which we shall denote with a prime ($'$).
The interactions of the mirror particles have the same form
as the standard particles, so that the Lagrangian is essentially doubled:
 \begin{eqnarray}
 {\cal L} = {\cal L}_{SM} (e, d, u, \gamma, ...) + {\cal L}_{SM} (e', d', u', \gamma', ...)
 \end{eqnarray}
 At this stage, the two sectors are essentially decoupled from each other except
via gravity (although we will 
 discuss the possible ways in which the two sectors can interact with each other in a moment). 
 In such a theory,
 the mirror baryons are naturally dark, stable and massive and are therefore, a priori, 
 excellent candidates for dark matter. The theory exhibits a gauge symmetry
which is $G_{SM} \otimes G_{SM}$ (where $G_{SM} = SU(3)_c \otimes SU(2)_L \otimes U(1)_Y$
is the standard model gauge symmetry).
 
 One can define a discrete symmetry interchanging ordinary and mirror particles, which can 
 be interpreted as space-time parity symmetry ($x \to -x$) if the roles of left and right chiral fields
 are interchanged in the mirror sector. Because of this geometical interpretation, one cannot
 regard this discrete symmetry as ad hoc in any sense.
 
 An obvious question is: can ordinary and mirror particles interact with each other non-gravitationally?
 The answer is YES - but only two terms are consistent with renormalizability and symmetry\cite{flv}:
 \begin{eqnarray}
 {\cal L}_{mix} = \frac{\epsilon}{2} F^{\mu \nu} F'_{\mu \nu} + \lambda \phi^{\dagger}\phi \phi'^{\dagger} \phi' \ ,
 \end{eqnarray}
 where $F_{\mu \nu}$ ($F'_{\mu \nu}$) is the ordinary (mirror) $U(1)$ gauge boson field strength tensor and 
 $\phi$ ($\phi'$) is the electroweak Higgs (mirror Higgs) field.
 These two terms are very important, because they lead to ways to experimentally test the idea.

 With the above Higgs - mirror Higgs quartic coupling term included, the full Higgs potential
of the model has three parameters. Minimizing this potential, 
one finds that there are two possible vacuum solutions (with each solution holding for a range of parameters): 
$\langle \phi \rangle = \langle \phi' \rangle \simeq 174$ GeV (unbroken mirror symmetry) and $\langle \phi \rangle \simeq 174$ GeV, $\langle \phi' \rangle = 0$ (spontaneously broken mirror symmetry\footnote{
Mirror QCD effects eventually break $SU(2)\times U(1)$ in the mirror sector
leading to a small, but non-zero VEV for $\phi'$ in the spontaneously
broken case. See Ref.\cite{bro} for details.}). While both vacuum solutions are phenomenologically viable, we shall
henceforth assume that the mirror symmetry is unbroken, because that case seems more interesting from
a dark matter perspective.
In the unbroken mirror symmetry case the mass and interactions of the mirror particles 
are exactly the same as the ordinary particles (except for the interchange of left and right).
 
 Is mirror matter too much like ordinary matter to account for the non-baryonic dark matter in the Universe?
 After all, ordinary and dark matter have some different properties:
 \begin{itemize}
 \item Dark matter is (roughly) spherically distributed in spiral galaxies, which is in sharp 
 contrast to ordinary matter which has collapsed onto the disk.
 
 \item $\Omega_{dark} \neq \Omega_b$ but $\Omega_{dark} \approx 5\Omega_b$.
 
 \item Big Bang Nucleosynthesis (BBN) works very well without any extra energy density
 from a mirror sector.
 
 \item
 Large scale structure formation should begin prior to ordinary photon decoupling.
 
 \end{itemize}
 \noindent
 Clearly there is no `macroscopic' symmetry. But this doesn't preclude the possibility of exactly
 symmetric microscopic physics. Why? Because the initial conditions in the Universe 
 might be different in the two sectors. In particular, if in the early Universe, 
 the temperature of the mirror particles ($T'$) were significantly less than the ordinary 
 particles ($T$) then:
 \begin{itemize}
 \item Ordinary BBN is not significantly modified provided $T' \stackrel{<}{\sim} 0.5 T$.
 
 \item $\Omega_{dark} \neq \Omega_b$ since baryogenesis mechanisms typically depend
 on temperature\footnote{The fact that $\Omega_{dark} \neq \Omega_b$ but $\Omega_{dark} \sim \Omega_b$
 is suggestive of some similarity between the ordinary and dark matter particle properties, which
 might be explained within the mirror dark matter context by having exactly symmetric
 microscopic physics and asymmetric temperatures. For some specific models in this
 direction, see ref.\cite{s1,s2}\ .}.
 
 \item
 Structure formation in the mirror sector can start before ordinary photon decoupling because
 mirror photon decoupling occurs earlier if $T' < T$\cite{b1}. Detailed studies\cite{b1b} find that
for $T' \stackrel{<}{\sim} 0.2T$ successful large scale structure follows. This dark matter candidate is also
 nicely consistent with CMB measurements\cite{b2}.
 
 \item
 Furthermore, BBN in the mirror sector is quite different since mirror BBN occurs earlier
 if $T' < T$. In fact, because of the larger expansion rate at earlier times we would expect
 that the $He'/H'$ ratio be much larger than the ratio of $He/H$ in the Universe. 
 This would change the way
 mirror matter evolves on short scales c.f. ordinary matter. 
 Maybe this can explain why mirror matter hasn't yet collapsed onto the disk\cite{fv}.
 
 \end{itemize}
 
\noindent
 Ok, so mirror matter can plausibly explain the non-baryonic dark matter inferred to exist
 in the Universe.
 Can it really be experimentally tested though?
 
 The Higgs mixing term will impact on the properties of the standard model Higgs\cite{flv2,sasha}.
 This may be tested if a scalar is found in experiments, e.g. at the forthcoming LHC experiment.
 More interesting, at the moment, is the $\epsilon F^{\mu \nu}F'_{\mu \nu}$ term. This interaction
 leads to kinetic mixing of the ordinary photon with the mirror photon, which in turn leads
 to orthopositronium - mirror orthopositronium oscillations\cite{gl} (see also \cite{fg}).
 Null results of current experiments imply\cite{update} $\epsilon < 5\times 10^{-7}$.
 Another consequence of the $\epsilon F^{\mu \nu}F'_{\mu \nu}$ term is that it will lead to elastic (Rutherford) scattering of mirror baryons off ordinary baryons, since the  mirror proton effectively couples to ordinary photons with electric charge $\epsilon e$.
 This means that conventional dark matter detection experiments currently searching
 for WIMPs can also search for mirror dark matter!\cite{my1}
 The DAMA/NaI experiment already claims direct detection of dark matter\cite{dama}. Can 
 mirror dark matter explain that experiment?
 
 The interaction rate in an experiment such as DAMA/NaI has the general form:
 \begin{eqnarray}
 \frac{dR}{dE_R} = \sum_{A'} N_T n_{A'} \int^{\infty}_{v'_{min} (E_R)} \frac{d\sigma}{dE_R} 
 \frac{f(v', v_E)}{k} |v'| d^3 v'
 \label{4a}
 \end{eqnarray}
 where $N_T$ is the number of target atoms per kg of detector, $n_{A'}$
is the galactic halo number
 density of dark matter particles labeled as $A'$. We include a sum allowing for
 more than one type of dark matter particle. In the above equation $f(v',v_E)/k$ is the
 velocity distribution of the dark matter particles, $A'$, and $v_E$ is the Earth's velocity relative
 to the galaxy. Also, $v'_{min} (E_R)$ is the minimum velocity for which a dark matter
 particle of mass $M_{A'}$ impacting on a target atom of mass $M_A$ can produce a recoil
 of energy $E_R$ for the target atom. This minimum velocity satisfies the kinematic 
 relation: 
 \begin{eqnarray}
 v'_{min}(E_R) = \sqrt{ \frac{(M_A + M_{A'})^2 E_R}{2 M_A M^2_{A'} }}
 \end{eqnarray}
 
 The DAMA experiment eliminates the background by using the annual modulation signature. The idea\cite{idea} is very simple.
 The rate, Eq.\ref{4a}, must vary periodically since it depends on the Earth's velocity, $v_E$, which 
modulates due to the Earth's motion around the Sun. That is,
 \begin{eqnarray}
 R(v_E) = R (v_{\odot}) + \left( \frac{\partial R}{ \partial v_E}\right)_{v_{\odot}} \Delta v_E \cos \omega (t-t_0)
 \end{eqnarray}
 where $\Delta v_E \simeq 15$ km/s, $\omega \equiv  2\pi/T$ ($T = 1$ year) and $t_0 = 152.5$ days (from
astronomical data).
 The phase and period are both predicted! This gives a strong systematic check on their results.
 Such an annual modulation was found\cite{dama} at the $6.3\sigma$ Confidence level, with $T, t_0$ measured to be:
 \begin{eqnarray}
 T &=& 1.00 \pm 0.01 \ {\rm year} \nonumber \\
 t_0 &=& 140 \pm 22 \ {\rm days}
 \end{eqnarray}
 Clearly, both the period and phase are consistent with the theoretical expectations of halo dark matter.

 The signal occurs in a definite low energy range from 6 keVee down to the experimental threshold
 of 2 keVee\footnote{The unit, keVee is the so-called electron equivalent energy,
which is the energy of an event if it were due to an electron recoil. The actual nuclear recoil 
energy (in keV) is given by: ${\rm keVee}/q$, where $q$ is the quenching factor ($q_I \simeq 0.09$ and
$q_{Na} \simeq 0.30$).}. No annual modulation was found for $E_R > 6$ keVee. 
 Given that the mean velocity of halo
 dark matter particles relative to the Earth is of order the local rotational velocity  
($\sim 300$ km/s), this suggests a mass for the (cold) dark matter
 particles roughly of order 20 GeV, since:
 \begin{eqnarray}
 E &=& \frac{1}{2} m v^2 
  \simeq   \frac{m}{20 \ {\rm GeV}} \left( \frac{v}{300 \ {\rm km/s}}\right)^2 \ 10 \ {\rm keV.}
  \end{eqnarray} 
 Dark matter particles with mass larger than about 60 GeV would give
 a signal above the 6 keVee region (no such signal was observed in the DAMA experiment).
 On the other hand, dark matter particles with mass less than about 5 GeV do not have enough energy
 to produce a signal in the 4-6 keVee energy region -  which would be contrary to the DAMA results.
 Importantly, the mass region sensitive to the DAMA experiment coincides with that
 predicted by mirror dark matter, since mirror dark matter predictes a spectrum of dark matter
 elements ranging in mass from hydrogen to iron. That is, with  mass 
$ {\rm GeV} \stackrel{<}{\sim} M_{A'} \stackrel{<}{\sim} 55\ {\rm GeV}$.
 A detailed analysis\cite{my1} confirms that mirror dark matter can fit the DAMA experimental data and the required value for $\epsilon$ is $\epsilon \sim 10^{-9}$. This fit to the
annual modulation signal is given in {\bf figure 1}.

\begin{figure}
\psfig{file=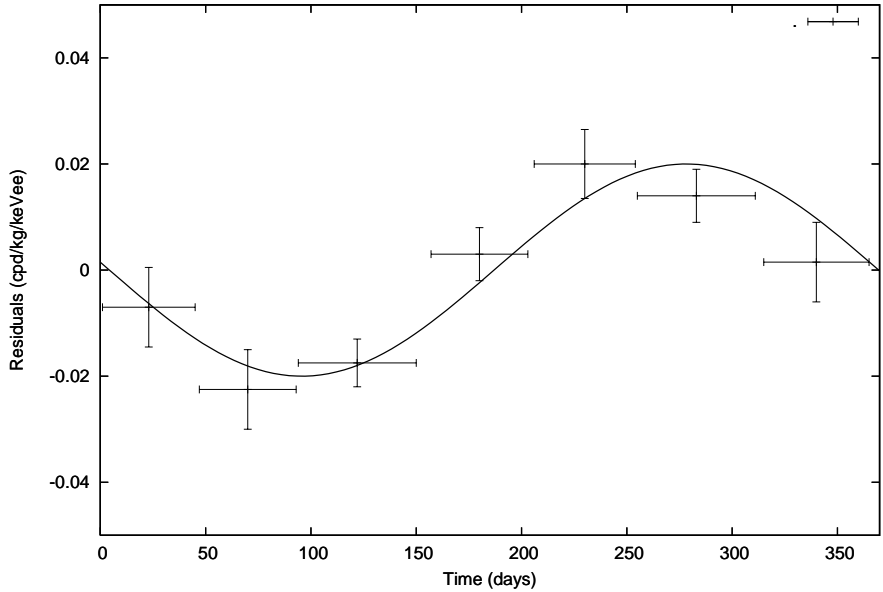,angle=270,width=12.1cm}

\caption{DAMA/NaI annual modulation signal (taking data from ref.\cite{dama}) together with the mirror matter prediction. Note
that the initial time in this figure is August 7th.}
\end{figure}

Interestingly, a mirror
sector interacting with the ordinary particles with $\epsilon \sim 10^{-9}$ has
many other interesting applications (see e.g. ref.\cite{i1,i2}). It also consistent with the Laboratory
(orthopositronium) bound as well as BBN constraints\cite{cg}.

What about the null results of the other direct detection experiments, such as the CDMS, Zeplin, Edelweiss experiments?
 For any model which explains the DAMA/NaI annual modulation signal, the corresponding rate
for the other direct detection experiments can be predicted.
 These null results do seem to disfavour the WIMP interpretation of the DAMA experiment. However
 it turns out that they do not, at present, disfavour the mirror dark matter
 interpretation. Why? because these other experiments are typically all higher threshold experiments with
 heavier target elements than Na (which, in the mirror matter interpretation, dominates
 the DAMA/NaI signal)
 and mirror dark matter has three key features which make it less sensitive (than WIMPs) to higher
 threshold experiments.
 \begin{itemize}
 \item Mirror dark matter is relatively light $M_H \le M_{A'} \le M_{Fe}$.
 
 \item The Rutherford cross section has the form:
 \begin{eqnarray}
 \frac{d\sigma}{dE_R} \propto \frac{1}{E_R^2}  \nonumber
 \end{eqnarray}
 while for WIMPs it is $E_R$ independent (excepting the energy dependence of the
form factors).
 
 \item Mirror particles interact with each other. This implies that the Halo particles are
 in local thermodynamic equilibrium, 
 so that e.g. $T = \frac{1}{2} M_{H'} \overline{v_{H'}^2} = \frac{1}{2} M_{O'}
 \overline{v_{O'}^2}$ ($\approx$ 300 eV assuming the standard assumptions of 
 an isothermal halo in hydrostatic equilibrium\cite{review}). Thus heavier elements have smaller mean velocities.
 
 \end{itemize}

 To summarize, having a mirror sector is a simple way to explain the inferred dark matter
 of the Universe. There is experimental support for this particular
 dark matter hypothesis, coming from the positive DAMA annual modulation signal. We must
 await future experiments to see if this explanation is the correct hypothesis.

 \vskip 1cm
 \noindent
 {\bf Acknowledgements:} This work was supported by the Australian Research Council.

 \end{document}